\documentstyle[secnumeq]{FBSart}
\newenvironment{Eqnarray}{\arraycolsep 0.14em\begin{eqnarray}}{\end{eqnarray}}
\def\s{\rule{0pt}{10pt}}
\def\t{\rule{0pt}{8pt}}

\title{Low Energy Proton-Deuteron Scattering in Configuration Space}
\author{C. R. Chen\instnr{1}, J. L. Friar\instnr{2}, G. L. Payne\instnr{3}}
\instlist{Department of Physics, Texas A\&M University, College Station,
TX 77843, USA \and
Theoretical Division, Los Alamos National Laboratory, Los Alamos, NM 87545, USA
\and Department of Physics and Astronomy, University of Iowa, Iowa City,
IA 52242, USA}

\runningauthor{C. R. Chen}
\runningtitle{Low Energy Proton-Deuteron Scattering in Configuration Space}

\begin{document}

\maketitle

\begin{abstract}

A new method for solving the configuration-space Faddeev equations for elastic
{\it p-d} scattering below the deuteron-breakup threshold is described.
Numerical solutions that demonstrate the convergence and accuracy of the
method are given. The number of channels and the value of the matching radius
required to obtain an accurate solution are also investigated.
These calculations demonstrate that this method can efficiently solve
the large matrix equations required for the three-body scattering problem.

\end{abstract}

\section{Introduction}

The three-nucleon system is a useful tool for investigating our understanding
of the nuclear force. Since it is possible to obtain accurate numerical
solutions for the three-body system, detailed comparisons of the model
Hamiltonian with the experimental data can be performed. These comparisons
provide stringent tests of the two-body interactions obtained by fitting the
two-body data, and give a means of studying the three-nucleon force.
Accurate bound-state calculations \cite{triton} have shown that realistic
two-body interactions yield binding energies for the triton and $^3{\rm He}$
that are less than the experimental values, and one must add a three-body
interaction to obtain the correct value. The bound-state system unfortunately
provides limited information about the three-body force. Most of the
bound-state properties scale with the binding energy \cite{scaling}. That is,
different interactions that give the same trinucleon binding energy predict
nearly the same charge radii, asymptotic normalization constants, magnetic
moments, etc.

Initially it was believed that the nucleon-deuteron scattering problem offered
an opportunity to explore our understanding of the three-nucleon force; however,
it has been shown \cite{report} that most of the scattering data can be
reproduced at a good level with only two-body interactions. For low-energy
scattering the effects of the three-nucleon force are usually small; there are
nevertheless some discrepancies (such as the nucleon analyzing power $A_y$ and
the deuteron analyzing power $iT_{11}$) that could be sensitive to these
interactions. Since much of the experimental data is for {\it p-d} scattering
at low energies, where the Coulomb interaction cannot be neglected, it is
necessary to solve the scattering equations for this case as well as for the
{\it n-d} case.

The Pisa group \cite{pisa} has shown that it is possible to obtain accurate
solutions for the {\it p-d} scattering equations using the Pair-correlated
Hyperspherical Harmonic (PHH) basis to expand the wave function, and solving
for the S-matrix by using the complex form of the Kohn variational principle.
In this paper we describe an alternate method for solving the
configuration-space equations. While the method is similar to the one used in
our previous calculations of the scattering length \cite{chen}, a new iterative
technique for solving the large matrix equations is presented. In addition to
demonstrating the convergence and accuracy of this technique, we provide a
detailed study of the number of angular-momentum and spin-isospin states
required to obtain an accurate solution. Since the question of how large a
matching radius is required in configuration space to obtain an accurate
solution has been a concern about this method, we also investigate the
convergence of the method as a function of this parameter. While the present
work is limited to scattering below the three-body breakup threshold, the
method can be extended to higher energies using the boundary conditions
discussed in Ref.\ \cite{chapter}.

The new numerical method is described in the next section, and the convergence
of the procedure is illustrated in Sect.\ 3.

\section{Numerical Methods}

We use the Jacobi coordinates
\begin{Eqnarray}
{\bf x}_i & \equiv & {\bf r}_j - {\bf r}_k \label{xdef} \\
\noalign{\hbox{and}}
{\bf y}_i & \equiv & \frac{1}{2} ({\bf r}_j + {\bf r}_k) - {\bf r}_i\,,
         \label{ydef}
\end{Eqnarray}
\hspace*{-0.28em}where $i$, $j$, and $k$ imply cyclic permutation.
The Hamiltonian in the center-of-mass frame is
\begin{equation}
H = T + V({\bf x}_1) + V({\bf x}_2) + V({\bf x}_3) + V_C\,, \label{ham}
\end{equation}
where
\begin{equation}
V_C = \sum_{i=1}^3 \frac{e^2}{x_i}\,\frac{[1+\tau_z(j)][1+\tau_z(k)]}{4}
      \label{coulpot}
\end{equation}
is the sum over the two-body Coulomb interactions between the pairs and $V({\bf
x}_i)$ is the nucleon-nucleon potential between the pair $j$ and $k$. For an
incident nucleon with kinetic energy $E_{\rm lab}$ in the laboratory frame, the
total energy in the center-of-mass frame is the sum of the deuteron binding
energy and the kinetic energy of the incident particle in the center-of-mass
frame, $E_{\rm cm} = \frac{2}{3} E_{\rm lab}$. The corresponding wave numbers
are given by
\begin{equation}
E = E_{\rm d} + E_{\rm cm} = -\frac{\hbar^2 \kappa^2}{M} + \frac{3\hbar^2
    q^2}{4 M} \equiv -\frac{\hbar^2 {\mathcal{K}}^2}{M}\,, \label{energy}
\end{equation}
where $M$ is the nucleon mass, $\kappa$ is the deuteron bound-state wave number,
and $q$ is the wave number of the nucleon or deuteron in the center-of-mass
frame.

Writing the total wave function $\Psi$ as the sum of the three Faddeev
amplitudes
\begin{equation}
\Psi = \Psi_1({\bf x}_1,{\bf y}_1) + \Psi_2({\bf x}_2,{\bf y}_2) +
       \Psi_3({\bf x}_3,{\bf y}_3)\,, \label{totpsi}
\end{equation}
the Schr\"odinger equation can be decomposed into the three Faddeev equations
\begin{equation}
[T + V({\bf x}_i) + V_C -E] \Psi_i = -V({\bf x}_i) (\Psi_j + \Psi_k) \,.
\end{equation}
Adding these three equations gives the Schr\"odinger equation.
For three identical particles the $\Psi_i$ all have the same functional form,
and we need to solve only the $i=1$ equation. To solve this equation, we make a
partial-wave expansion of the Faddeev amplitudes using the $j-J$ coupling
scheme. We write
\begin{equation}
\Psi_i({\bf x}_i,{\bf y}_i) = \sum_\alpha \frac{\psi_\alpha(x_i,y_i)}{x_i y_i}
                    \vert \alpha \rangle_{\!i \s} \,,
\end{equation}
where
\begin{equation}
\vert \alpha \rangle_{\!i \s}= \vert \left[ (l_\alpha,s_\alpha)j_\alpha,
       (L_\alpha,S_\alpha)J_\alpha\right]JM;(t_\alpha,T_\alpha)TM_T \rangle
       \label{channel}
\end{equation}
is the orbital angular momentum and spin-isospin state function for the
different channels, and $i$ indicates that the order of coupling is
$[(j,k),i]$ for cyclic values of $i$, $j$, and $k$. The relative orbital
angular momentum of the $j$-$k$ pair is $l_\alpha$, the spin of the pair is
$s_\alpha$, and the total angular momentum of the pair is $j_\alpha$. The
orbital angular momentum of particle $i$ relative to the center-of-mass of the
$j$-$k$ pair is $L_\alpha$, $S_\alpha$ is the spin of particle $i$, and
$J_\alpha$ is the total angular momentum of the particle. Finally, $J$ is the
total angular momentum of the system. For the isospin function, $t_\alpha$ is
the isospin of the $j$-$k$ pair, $T_\alpha$ is the isospin of particle $i$, and
$T$ is total isospin of the system. For nucleons $S_\alpha$ and $T_\alpha$ are
both $1/2$.

The Faddeev equation for $\Psi_1({\bf x}_1,{\bf y}_1)$ can be reduced to a set
of coupled partial differential equations for the reduced amplitudes
$\psi_\alpha$ by taking the inner product with the state functions
$\vert \alpha \rangle_{\!1 \t}$. After multiplying the equations by $M/\hbar^2$
and transforming to the hyperspherical coordinates defined by
\begin{Eqnarray}
x &=& \rho \cos \theta \\
\noalign{\hbox{and}}
y &=& \frac{\sqrt 3}{2} \rho \sin \theta \,,
\end{Eqnarray}
\hspace*{-0.28em}the resulting equations are
\begin{Eqnarray}
\left(\Delta_\alpha \right.&-& \left. {\mathcal{K}}^2 \right)
   \psi_\alpha(\rho,\theta) - \sum_{\alpha^\prime} \left(v_{\alpha
   \alpha^\prime} + v_{\alpha\alpha^\prime}^C \right)\psi_{\alpha^\prime}
   (\rho,\theta) \nonumber \\
   & & \qquad -\sum_{\alpha^\prime} v_{\alpha \alpha^\prime}
   \sum_{\alpha^{\prime \prime}}
   \int_{\theta^-}^{\theta^+} \tilde K_{\alpha^\prime \alpha^{\prime \prime}}(\theta,
   \theta^\prime)\psi_{\alpha^{\prime \prime}}(\rho, \theta^\prime)\,d
   \theta^\prime = 0, \label{diffeqn1}
\end{Eqnarray}
\hspace*{-0.28em}where
\begin{Eqnarray}
\Delta_\alpha &=& \frac{\partial^2}{\partial \rho^2} + \frac{1}{\rho}
      \frac{\partial}{\partial \rho} + \frac{1}{\rho^2}\frac{\partial^2}
      {\partial \theta^2} -\frac{l_\alpha(l_\alpha+1)}{\rho^2 \cos^2\theta}
      - \frac{L_\alpha (L_\alpha+1)}{\rho^2 \sin^2\theta}\,, \nonumber \\
v_{\alpha \alpha^\prime} &=& \frac{M}{\hbar^2}\,_{1 \s\mkern -5.5mu} \langle
      \alpha \vert V({\bf x}_1) \vert \alpha^\prime \rangle_{\!1 \s} \,,\\
v_{\alpha\alpha^\prime}^C &=&  \frac{M}{\hbar^2}\, _{1 \s\mkern -5.5mu}\langle
      \alpha \vert V^C(x_1,x_2,x_3) \vert \alpha^\prime \rangle_{\!1  \s} \,,
      \nonumber
\end{Eqnarray}
\hspace*{-0.28em}and
\begin{equation}
\sum_{\alpha^{\prime \prime}} \int_{\theta^-}^{\theta^+}
   \tilde K_{\alpha^\prime \alpha^{\prime \prime}}(\theta,\theta^\prime)
   \psi_{\alpha^{\prime \prime}} (\rho, \theta^\prime)\,d\theta^\prime
   = x_1 y_1\, _{1\s \mkern -5.5mu} \langle
   \alpha^\prime \vert \Psi_2({\bf x}_2,{\bf y}_2) + \Psi_3({\bf x}_3,{\bf y}_3)
   \rangle \,.
\end{equation}
For three equal-mass particles
the limits for the integration with respect to the hyperspherical angle
$\theta^\prime$ are given by
\begin{Eqnarray}
\theta^- &=& \left\vert \theta - \frac{\pi}{3} \right\vert \nonumber \\
\noalign{\hbox{and}}
\theta^+ &=& \frac{\pi}{2}-\left\vert \theta-\frac{\pi}{6}\right\vert \,,\nonumber
\end{Eqnarray}
\hspace*{-0.28em}where $0 \le \theta \le \pi/2$.

For the numerical solution of the coupled partial differential equations, we
choose a value, $\rho_{\rm max}$, for the maximum value of the $\rho$ variable,
and solve the equations for $0 \le \rho \le \rho_{\rm max}$. The equations have
a unique solution only if one specifies the boundary conditions on the closed
boundary of this region of the $\rho-\theta$ plane. Since the $\psi_\alpha$ are
reduced wave functions they must vanish as $x$ or $y$ goes to zero; thus, we
have the boundary conditions
\begin{equation}
\psi_\alpha(0,\theta) = \psi_\alpha(\rho,0) = \psi_\alpha(\rho,\pi/2) = 0
\end{equation}
for all channels. The boundary conditions for $\rho_{\rm max}$ depend on
whether the channel is open or closed. For energies below the
three-body-breakup threshold, the channels that correspond to elastic
scattering are the only open ones. The other channels, which are virtual
breakup channels, must decrease exponentially for large values of $\rho$. The
channels whose values for $l_\alpha$, $s_\alpha$, and $j_\alpha$ correspond to
the deuteron values are open, and they have the asymptotic form
\begin{equation}
\psi_\alpha(x_1,y_1) \,{\lower7pt\hbox{$\displaystyle\longrightarrow$}
    \atop\scriptstyle{y_1\rightarrow\infty}}\, \phi_\alpha(x_1,y_1) +
    \Omega_\alpha(x_1,y_1) \,,
\end{equation}
where $\phi_\alpha(x_1,y_1)$ is the component of the incident wave for the
channel $\alpha$ and $\Omega_\alpha(x_1,y_1)$ is the corresponding component of
the scattered wave. For {\it p-d} scattering $\phi_\alpha(x_1,y_1)$ is given by
\begin{equation}
\phi_\alpha(x_1,y_1) = F_{L_\alpha}\!(\eta,qy_1)\, u_{l_\alpha}(x_1) \,,
 \label{regcoul}
\end{equation}
where $F_{L_\alpha}\!(\eta,qy_1)$ is the regular Coulomb wave function with
$\eta = 2Me^2/3\hbar^2q$ and $u_{l_\alpha}$ is the $l_\alpha$ component of the
reduced deuteron bound-state wave function. For scattering energies below the
breakup threshold the K-matrix form of the scattering equations is real, and
using this form of the equations simplifies the numerical calculations since
the $\Omega_\alpha(x_1,y_1)$ are real functions. For a deuteron with
$j_\alpha = 1$ and a fixed value of the total angular momentum $J$ and parity
there are at most three open channels, and the equations must be solved for
incident waves in each of these channels. For an incident wave in channel
$\alpha^\prime$, the scattered wave in the K-matrix formulation has the form
\begin{equation}
\Omega_\alpha(x_1,y_1) = K_{\alpha \alpha^\prime}\, G\!_{L_\alpha}\!(\eta,
     qy_1) \,u_{l_\alpha}(x_1) \,,
\end{equation}
where $G\!_{L_\alpha}\!(\eta,qy_1)$ is the irregular Coulomb wave function.
Given the K-matrix one can determine the S-matrix using
\begin{equation}
S = \frac{1 + iK}{1 - iK} \,.
\end{equation}

For the numerical calculations we write the reduced Faddeev amplitudes in the
form
\begin{equation}
\psi_\alpha(x_1,y_1) = \delta_{\alpha\alpha^\prime}\phi_\alpha(x_1,y_1)
                      + \chi_\alpha(\rho,\theta) \,,   \label{reducedamp}
\end{equation}
for the case with an incident deuteron in the channels with a fixed value of
$J_{\alpha^\prime}$. We define the function $F_\alpha(\rho,\theta)$ (which is not
the regular Coulomb function used in Eq. (\ref{regcoul})) by writing
\begin{equation}
\chi_\alpha(x_1,y_1) = F_\alpha(\rho,\theta) e^{-{\mathcal{K}}\rho}
   \label{closed}
\end{equation}
for the closed channels, and
\begin{equation}
\chi_\alpha(x_1,y_1) = \frac{F_\alpha(\rho,\theta)}{x_1} u_{l_\alpha}
                    \label{open}
\end{equation}
for the open channels, where the deuteron wave function has been factored out
of the elastic channels to make $F_\alpha(\rho,\theta)$ a smoother function.
The $x_1$ in the denominator of Eq. (\ref{open}) is included to force
$F_\alpha(\rho,\theta)$ to vanish when $x_1$ is zero. The boundary condition
for closed channels is that $\chi_\alpha(x_1,y_1)$ go to zero for large values
of $\rho$, which can be implemented by the condition
\begin{equation}
\frac{\partial F_\alpha}{\partial \rho}\Bigg\vert_{\rho=\rho_{\rm max}}
           = 0 \,;
\end{equation}
that is, $F_\alpha(\rho,\theta)$ is a constant for large values of $\rho$.
For the open channels
\begin{equation}
\frac{F_\alpha(x_1,y_1)}{x_1} \,{\lower7pt\hbox{$\displaystyle\longrightarrow$}
    \atop\scriptstyle{y_1\rightarrow\infty}}\, K_{\alpha \alpha^\prime}\,
    G\!_{L_\alpha}\!(\eta,qy_1) \,, \label{kmat}
\end{equation}
which is implemented by the boundary condition
\begin{equation}
\frac{\partial F_\alpha}{\partial \rho}\Bigg\vert_{\rho=\rho_{\rm max}}
        = \cos\theta \frac{F_\alpha}{x_1} + \frac{\sqrt 3}{2} \sin\theta
         \frac{F_\alpha}{G\!_{L_\alpha}} \frac{d G\!_{L_\alpha}}{d y_1} \,.
\end{equation}

To solve the equations for a given value of the total angular momentum, $J$,
and parity, we truncate the number of channels by choosing a maximum value for
the angular momentum, $j_\alpha$, of the interacting pair and solve the
differential equation for an incident wave in each of the open channels.
Substituting Eq.\ (\ref{reducedamp}) into Eq\ (\ref{diffeqn1}) gives
\begin{Eqnarray}
\left(\Delta_\alpha \right.&-& \left. {\mathcal{K}}^2 \right)
\chi_\alpha(\rho,\theta) - \sum_{\alpha^\prime} \left(v_{\alpha \alpha^\prime}
   + v_{\alpha\alpha^\prime}^C \right)\chi_{\alpha^\prime}(\rho,\theta)
   \nonumber \\
   & & \quad\quad -\sum_{\alpha^\prime} v_{\alpha \alpha^\prime}
   \sum_{\alpha^{\prime \prime}}
   \int_{\theta^-}^{\theta^+} \tilde K_{\alpha^\prime \alpha^{\prime \prime}}
   (\theta,\theta^\prime)\chi_{\alpha^{\prime \prime}}(\rho, \theta^\prime)\,d
   \theta^\prime \label{diffeqn2} \\
        &= &  \delta_{\alpha\beta} \sum_{\beta^\prime}
          (v_{\beta\beta^\prime}^C - \omega \delta_{\beta\beta^\prime})
          \phi_{\beta^\prime} (\rho,\theta)
   +\sum_{\alpha^\prime} v_{\alpha \alpha^\prime}
   \sum_{\beta^\prime} \int_{\theta^-}^{\theta^+}
      \tilde K_{\alpha^\prime \beta^\prime}(\theta,\theta^\prime)
      \phi_{\beta^\prime}(\rho, \theta^\prime)\,d\theta^\prime\,,
                     \nonumber
\end{Eqnarray}
\hspace*{-0.28em}where the sum over $\beta^\prime$ is over the channels that
have an incident deuteron bound state in the asymptotic region, and $\omega =
(M/\hbar^2)e^2/y_1$. Finally, substitution of Eqs.\ (\ref{closed}) and
(\ref{open}) into (\ref{diffeqn2}) gives a set of coupled partial differential
equations for the $F_\alpha(\rho,\theta)$. To solve these equations we expand
$F_\alpha(\rho,\theta)$ in a complete set of basis functions that is the tensor
product of Hermite splines for the $\rho$ and $\theta$ variables. We write
\begin{equation}
F_\alpha(\rho,\theta) = \sum_{i=1}^{N_\rho}\sum_{j=1}^{N_\theta} a_{ij}^\alpha
                         s_i(\rho)s_j(\theta) \,, \label{splinex}
\end{equation}
and use the orthogonal collocation method to determine the $a_{ij}^\alpha$; that
is, we require (\ref{splinex}) to satisfy the differential equation at the
collocation points $(\rho_l,\theta_m)$ for $l=1,\ldots,N_\rho$ and
$m = 1,\ldots,N_\theta$. This gives a matrix equation whose columns are labeled
by the values of $i$, $j$, and $\alpha$, and the rows are labeled by the
values of $l$, $m$, and $\alpha^\prime$. For an accurate solution to the
equations the number of the expansion coefficients, $a_{ij}^\alpha$, can exceed
several hundred thousand, and the resulting matrix equation is too large to
invert directly. From Eq.\ (\ref{diffeqn2}) and the local nature of the
splines, one can see that the matrix will have a block-diagonal structure.
Since the tensor component of the nuclear force couples at most two channels,
the matrix for the first two terms on the left-hand side of Eq.
(\ref{diffeqn2}) will have a much smaller bandwidth than the third term, which
couples all of the channels. Thus we write the matrix equation in the form
\begin{equation}
(A-B)a = b \,, \label{matrix1}
\end{equation}
where the matrix $A$ can be inverted by standard methods, while the matrix $B$
with a much larger bandwidth requires the use of an iterative procedure.

Since standard iterative methods such as the Lanczos algorithm or Pad\'e
approximants required many iterations to converge to an accurate solution for
the matrix equation, we derive a new algorithm to solve the matrix equation.
Rewriting (\ref{matrix1}) in the form
\begin{equation}
\left( 1 - B A^{-1} \right) \left(Aa\right) = b \,, \label{aaeqn}
\end{equation}
we find an approximate solution for $Aa$ using a set of orthogonal basis
vectors constructed by the Gram-Schmidt procedure. Starting with $u_0=
b/\sqrt{b^T b}$, we iterate Eq.\ (\ref{aaeqn}) to generate the basis vectors.
Given a set of basis vectors, $u_i$ for $i=0,\ldots,N_i-1$, a new basis vector
is constructed by first generating the vector
\begin{equation}
v_{N_i} = \left( 1 - B A^{-1} \right) u_{N_i-1} \,, \label{veqn}
\end{equation}
and then using the Gram-Schmidt procedure to generate a $u_{N_i}$ which is
normalized and orthogonal to the other basis vectors.
After $N_i$ iterations, the basis vectors $u_i$ are used to find an approximate
solution for $Aa$ by writing
\begin{equation}
Aa \approx \sum_{i=0}^{N_i-1} c_i u_i \,. \label{approx}
\end{equation}
Substituting this approximation into Eq.\ (\ref{aaeqn}) and using Eq.\
(\ref{veqn}), gives
\begin{equation}
\sum_{i=0}^{N_i-1} c_i v_{i+1} = b \,.
\end{equation}
Taking the inner product with $u_i$ gives the matrix equation
\begin{equation}
\sum_{j=0}^{N_i-1} u_i^T v_{j+1} c_j = u_i^T b \,,
\end{equation}
which can be solved for the $c_i$.
Multiplying the approximation for $Aa$ by $A^{-1}$ gives the approximate
solution for $a$. Using this approximate solution, the $F_\alpha(\rho,\theta)$
can be evaluated using Eq.\ (\ref{splinex}), and the values of $K_{\alpha,
\alpha^\prime}$ can be determined using Eq.\ (\ref{kmat}). The procedure is
repeated until a solution of the desired accuracy is found. Normally 10-20
iterations are required to obtain an accurate solution.

\section{Numerical Results}

To demonstrate the convergence of the iterative method, we consider the case
of {\it p-d} scattering for the complete AV18 two-body potential \cite{AV18}
with $\hbar^2/M = 41.47108\, {\rm MeV\, fm^2}$. The K-matrix elements are
for the $J$-$j$ channel states defined in Eq.\ (\ref{channel}), which are
different than the channel states used for the {\it n-d} \cite{nd-benchmark}
and {\it p-d} \cite{pd-benchmark} benchmark calculations. Our results for
the K-matrices can be converted to the channel scheme using Eq.\ (2.26) in
Ref.\ \cite{nd-benchmark}.
The incident state has $T = 1/2$, and we neglect
the isospin mixing, which has been shown to be small for low-energy scattering
\cite{pdcalc}. Thus, for a given value of the total angular momentum, $J$, and
parity, the possible channel states in Eq.\ (\ref{channel}) are determined by
the values of $j_\alpha$ and $J_\alpha$. For our calculations the number of
channel states is determined by keeping all states with values of $j_\alpha$ up
to $j_{\rm max}$. In addition, the equations must be solved numerically in the
region $0\le \rho \le \rho_{\rm max}$, where $\rho_{\rm max}$ must be large
enough that the asymptotic boundary conditions are an accurate approximation to
the scattered wave function. It is shown below that using $j_{\rm max}=10$ and
$\rho_{\rm max}=90\,{\rm fm}$ gives accurate numerical solutions. To illustrate
the convergence of the iterative solution to the numerical equations, we show
in Table \ref{TAB1} the values for the K-matrix elements for the $J = 1/2^+$
scattering at $E_{\rm lab}=1.0\,{\rm MeV}$. For this case the indices 1 and
2 refer to the states with $(L_\alpha,J_\alpha)$ equal to $(0,1/2)$ and
$(2,3/2)$, respectively. The values for the $K_{ij}$ are
calculated using the approximate solution obtained from Eq. (\ref{approx}).

\begin{table}[htb]
\beforetab
\begin{tabular}{ccccc}
\firsthline
 $i$ & $K_{11}\times 10^1$ &
   $K_{12}\times 10^3$ & $K_{21}\times 10^3$ & $K_{22}\times 10^2$ \\
\midhline
        3 & -1.3847 & -9.2557 & -2.5439 & -1.3052  \\
        4 & -2.7632 & -2.5891 & -2.7734 & -1.2797  \\
        5 & -2.4279 & -3.8322 & -3.5321 & -1.3369  \\
        6 & -2.1698 & -4.3883 & -3.9463 & -1.3690  \\
        7 & -2.3100 & -3.8242 & -4.0234 & -1.3701  \\
        8 & -2.3003 & -4.0095 & -3.9951 & -1.3734  \\
        9 & -2.3082 & -3.9661 & -3.9724 & -1.3734  \\
       10 & -2.3092 & -3.9520 & -3.9597 & -1.3720  \\
       11 & -2.3073 & -3.9685 & -3.9645 & -1.3721  \\
       12 & -2.3076 & -3.9654 & -3.9653 & -1.3722  \\
       13 & -2.3076 & -3.9641 & -3.9649 & -1.3722  \\
       14 & -2.3077 & -3.9645 & -3.9650 & -1.3722  \\
       15 & -2.3076 & -3.9645 & -3.9650 & -1.3722  \\
       16 & -2.3076 & -3.9645 & -3.9650 & -1.3722  \\
\lasthline
\end{tabular}
\aftertab
\captionaftertab[]{\label{TAB1} Convergence of the iterative solution for the
K-matrix for $J = \frac{1}{2}^+$ {\it p-d} scattering at $E_{\rm lab}=1.0\,{\rm
MeV}$ evaluated using $\,j_{\rm max}=10$ and $\rho_{\rm max} = 90\,{\rm fm}$.}
\end{table}

To show the accuracy of the solution using a truncated set of channel states, we
list in Table \ref{TAB2} the values for the K-matrix elements calculated using
different values for $\,j_{\rm max}$ for two different energies. The value of
$N_c$ is the actual number of channel states. Since the correct K-matrix must
be symmetric, one test of the accuracy of the numerical solution is the
difference between $K_{ij}$ and $K_{ji}$. As the number of channels is
increased, this difference becomes smaller. From Table \ref{TAB2} one can see
that a value of $j_{\rm max}=10$ gives an accurate approximation for the
K-matrix.

\begin{table}[htb]
\beforetab
\begin{tabular}{crccccc}
\firsthline
  E(MeV) & $N_c$ & $j_{\rm max}$ & $K_{11}\times 10^1$ &
   $K_{12}\times 10^2$ & $K_{21}\times 10^2$ & $K_{22}\times 10^2$
   \\
\midhline
                      &   34 &   \phantom{1}4 &  -2.3131 & -.38871 & -.39693 & -1.3737  \\[0.5ex]
\lower 6pt \hbox{1.0} &   50 &   \phantom{1}6 &  -2.3081 & -.39424 & -.39688 & -1.3727  \\
                      &   66 &   \phantom{1}8 &  -2.3078 & -.39587 & -.39663 & -1.3724  \\[0.5ex]
                      &   82 &             10 &  -2.3076 & -.39645 & -.39650 & -1.3722  \\
\midhline
                      &   34 &   \phantom{1}4 &  -6.2352 & -1.0738 & -1.0859 & -6.2915  \\[0.5ex]
\lower 6pt \hbox{3.0} &   50 &   \phantom{1}6 &  -6.2261 & -1.0857 & -1.0872 & -6.2888  \\
                      &   66 &   \phantom{1}8 &  -6.2253 & -1.0882 & -1.0870 & -6.2878  \\[0.5ex]
                      &   82 &             10 &  -6.2250 & -1.0887 & -1.0870 & -6.2874  \\
\lasthline
\end{tabular}
\aftertab
\captionaftertab[]{\label{TAB2} Values of the K-matrix for $J = \frac{1}{2}^+$
{\it p-d} scattering evaluated using different numbers of basis states and
$\rho_{\rm max} = 90\,{\rm fm}$.}
\end{table}

Finally in Table \ref{TAB3} we show the results for different values of
$\rho_{\rm max}$. A value of $\rho_{\rm max}=90\,{\rm fm}$ is more than enough
to obtain an accurate solution to the scattering equations.

\begin{table}[htb]
\beforetab
\begin{tabular}{cccccc}
\firsthline
  $E_{\rm lab}$ & $\rho_{\rm max}$ & $K_{11}\times 10^1$ &
   $K_{12}\times 10^2$ & $K_{21}\times 10^2$ & $K_{22}\times 10^2$
   \\
\midhline
                      &   30 &  -2.2966 & -.37960 & -.39970 & -1.1755  \\[0.5ex]
\lower 6pt \hbox{1.0} &   50 &  -2.3076 & -.39681 & -.39681 & -1.3762  \\
                      &   70 &  -2.3076 & -.39652 & -.39657 & -1.3723  \\[0.5ex]
                      &   90 &  -2.3076 & -.39645 & -.39650 & -1.3722  \\
\midhline
                      &   30 &  -6.2440 & -1.0919 & -1.0972 & -6.5249  \\[0.5ex]
\lower 6pt \hbox{3.0} &   50 &  -6.2212 & -1.0887 & -1.0860 & -6.2867  \\
                      &   70 &  -6.2243 & -1.0888 & -1.0870 & -6.2878  \\[0.5ex]
                      &   90 &  -6.2250 & -1.0887 & -1.0870 & -6.2874  \\
\lasthline
\end{tabular}
\aftertab
\captionaftertab[]{\label{TAB3} Values of the K-matrix for $J = \frac{1}{2}^+$
{\it p-d} scattering as a function of the laboratory energy in MeV evaluated
using $\,j_{\rm max}=10$ and different values of $\rho_{\rm max}$ in fm.}
\end{table}

In Tables \ref{TAB4} and \ref{TAB5} we show the convergence of the $J=1/2^-$
state for a laboratory energy of $1.0\,{\rm MeV}$. For this case the indices 1
and 2 refer to the states with $(L_\alpha,J_\alpha)$ equal to $(1,1/2)$ and
$(1,3/2)$, respectively. The convergence is very similar to the $J=1/2^+$ case.

\begin{table}[htb]
\beforetab
\begin{tabular}{rccccc}
\firsthline
   $N_c$ & $j_{\rm max}$ & $K_{11}\times 10^1$ &
   $K_{12}\times 10^2$ & $K_{21}\times 10^2$ & $K_{22}\times 10^2$
   \\
\midhline
       34 &   \phantom{1}4 &  1.4461 & -6.0489 & -6.0613 & -4.0558  \\
       50 &   \phantom{1}6 &  1.4451 & -6.0495 & -6.0542 & -4.0581  \\
       66 &   \phantom{1}8 &  1.4446 & -6.0497 & -6.0517 & -4.0589  \\
       82 &             10 &  1.4445 & -6.0499 & -6.0508 & -4.0592  \\
\lasthline
\end{tabular}
\aftertab
\captionaftertab[]{\label{TAB4} Values of the K-matrix for $J = \frac{1}{2}^-$
{\it p-d} scattering at $E_{\rm lab}=1.0\,{\rm MeV}$ evaluated using different
numbers of basis states and $\rho_{\rm max} = 90\,{\rm fm}$.}
\end{table}

\begin{table}[htb]
\beforetab
\begin{tabular}{ccccc}
\firsthline
   $\rho_{\rm max}$ & $K_{11}\times 10^1$ &
   $K_{12}\times 10^2$ & $K_{21}\times 10^2$ & $K_{22}\times 10^2$
   \\
\midhline
       30 &  1.4277 & -5.9366 & -5.9524 & -3.9590  \\
       50 &  1.4448 & -6.0514 & -6.0521 & -4.0616  \\
       70 &  1.4445 & -6.0497 & -6.0505 & -4.0594  \\
       90 &  1.4445 & -6.0499 & -6.0508 & -4.0592  \\
\lasthline
\end{tabular}
\aftertab
\captionaftertab[]{\label{TAB5} Values of the K-matrix for $J = \frac{1}{2}^-$
{\it p-d} scattering at $E_{\rm lab}=1.0\,{\rm MeV}$ evaluated using $j_{\rm
max}=10$ and different values of $\rho_{\rm max}$.}
\end{table}

For $J$ greater than 1/2 there are three open channels, which increases the
magnitude of the numerical problem. In addition, there are more channels for
a given value of $j_{\rm max}$. To illustrate the convergence for higher
values of $J$, we consider the $J=5/2^+$ state for an incident laboratory energy
of $2.0\,{\rm MeV}$. For this case there are nine elements in the K-matrix. To
show the symmetry of the final K-matrix, we list the off-diagonal elements in a
separate table. The indices 1, 2 and 3 refer to the states with
$(L_\alpha,J_\alpha)$ equal to $(2,3/2)$, $(2,5/2)$ and $(4,7/2)$, respectively.
Table \ref{TAB6} illustrates the convergence of the diagonal elements as the
number of channels is increased, and Table \ref{TAB7} demonstrates the
convergence for the off-diagonal elements.

\begin{table}[htb]
\beforetab
\begin{tabular}{rcccc}
\firsthline
  $N_c$ & $j_{\rm max}$ & $K_{11}\times 10^3$ &
   $K_{22}\times 10^2$ & $K_{33}\times 10^3$
   \\
\midhline
       \phantom{1}82  & \phantom{1}4   & -8.44217 & -1.37566 & -1.40145 \\
                  130 & \phantom{1}6   & -8.45303 & -1.37703 & -1.40217 \\
                  178 & \phantom{1}8   & -8.45504 & -1.37725 & -1.40284 \\
                  226 &            10  & -8.45541 & -1.37725 & -1.40337 \\
\lasthline
\end{tabular}
\aftertab
\captionaftertab[]{\label{TAB6} Values of the diagonal elements of the K-matrix
for $J = \frac{5}{2}^+$ {\it p-d} scattering at $E_{\rm lab}=2.0\,{\rm MeV}$
evaluated using different numbers of basis states and $\rho_{\rm max} =
90\,{\rm fm}$.}
\end{table}

\begin{table}[htb]
\beforetab
\begin{tabular}{rccccccc}
\firsthline
   $N_c$ & $j_{\rm max}$ & $K_{12}\times 10^2$ & $K_{21}\times 10^2$ &
   $K_{13}\times 10^4$ & $K_{31}\times 10^4$ & $K_{23}\times 10^4$ &
   $K_{32}\times 10^4$  \\ \midhline
       \phantom{1}82  & \phantom{1}4 &  3.41079 &  3.40639 & -7.24681 & -7.40645 & -1.49972 & -1.54607 \\
                  130 & \phantom{1}6 &  3.40752 &  3.40611 & -7.33861 & -7.40684 & -1.52924 & -1.54380 \\
                  178 & \phantom{1}8 &  3.40673 &  3.40617 & -7.37622 & -7.40731 & -1.53775 & -1.54093 \\
                  226 &            10&  3.40650 &  3.40626 & -7.39359 & -7.40771 & -1.54067 & -1.53952 \\
\lasthline
\end{tabular}
\aftertab
\captionaftertab[]{\label{TAB7} Values of the off-diagonal elements of the
K-matrix for $J = \frac{5}{2}^+$ {\it p-d} scattering at $E_{\rm lab}=2.0\,{\rm
MeV}$ evaluated using different numbers of basis states and $\rho_{\rm max} =
90\,{\rm fm}$.}
\end{table}

Finally, in Tables \ref{TAB8} and \ref{TAB9} we show the convergence as the
value of $\rho_{\rm max}$ is increased. One can see that a value of $90\,{\rm
fm}$ is sufficient for an accurate solution.

\begin{table}[htb]
\beforetab
\begin{tabular}{cccc}
\firsthline
   $\rho_{\rm max}$ & $K_{11}\times 10^3$ &
   $K_{22}\times 10^2$ & $K_{33}\times 10^3$
   \\
\midhline
       30 & -8.47082 & -1.37961 & -0.83198 \\
       50 & -8.45433 & -1.37684 & -1.43016 \\
       70 & -8.45570 & -1.37721 & -1.40318 \\
       90 & -8.45541 & -1.37725 & -1.40337 \\
\lasthline
\end{tabular}
\aftertab
\captionaftertab[]{\label{TAB8} Values of the diagonal elements of the K-matrix
for $J = \frac{5}{2}^+$ {\it p-d} scattering at $E_{\rm lab}=2.0\,{\rm MeV}$
evaluated using $\,j_{\rm max}=10$ and different values of $\rho_{\rm max}$.}
\end{table}

\begin{table}[ht]
\beforetab
\begin{tabular}{ccccccc}
\firsthline
   $\rho_{\rm max}$ & $K_{12}\times 10^2$ & $K_{21}\times 10^2$ &
   $K_{13}\times 10^4$ & $K_{31}\times 10^4$ & $K_{23}\times 10^4$ &
   $K_{32}\times 10^4$  \\ \midhline
       30 &  3.41844 &  3.41881 & -6.24969 & -7.56114 & -1.30255 & -1.56836 \\
       50 &  3.40606 &  3.40580 & -7.43409 & -7.43557 & -1.55406 & -1.55135 \\
       70 &  3.40656 &  3.40632 & -7.40409 & -7.41772 & -1.54418 & -1.54286 \\
       90 &  3.40650 &  3.40626 & -7.39359 & -7.40771 & -1.54067 & -1.53952 \\
\lasthline
\end{tabular}
\aftertab
\captionaftertab[]{\label{TAB9} Values of the off-diagonal elements of the
K-matrix for $J = \frac{5}{2}^+$ {\it p-d} scattering at $E_{\rm lab}=2.0\,{\rm
MeV}$ evaluated using $j_{\rm max}=10$ and different values of $\rho_{\rm
max}$.}
\end{table}

\section{Conclusions}

A new numerical method for solving the large matrix equations in
configuration-space three-body scattering has been shown to be an efficient
procedure for obtaining accurate solutions for the realistic nucleon-deuteron
scattering problem. In addition, the convergence as a function of the number of
channels and the matching radius has been demonstrated.

\begin{acknowledge}
The work of J.\ L.\ F.\ was performed under the auspices of the U.\ S.\
Department of Energy. That of G.\ L.\ P.\ was supported in part by the U.\ S.\
Department of Energy.
\end{acknowledge}

\end{document}